# Landau Theory for States with local translation symmetry and the Phenomenological Potential with minimal electron-phonon interactions.

A.Ya. Braginsky

Research Institute of Physics, Southern Federal University, Rostov-on-Don

e-mail: a.braginsky@mail.ru

**Abstract**

In the paper, the inhomogeneous states with local translational symmetry are under investigation in the framework of the Landau theory. The phonon potential is defined as a compensating field of local representation of the subgroup of translations. The constructed phenomenological Ginzburg-Landau potential takes into account both electromagnetic and electron-phonon interactions explicitly. The suggested phenomenological model does not require doubling of charge to get the correct expression of current density in the Londons equations and is in agreement with the BCS model. A phenomenological description of sound based on the phonon potential is also proposed.

## 1. Introduction

Gauge models in the Landau theory with minimal interaction were introduced in the pioneer paper by Ginzburg and Landau [1] and later in the de Gennes model [2], to describe deformed SmA. Both models were borrowed from the field theory. In the present paper, we show that, vice versa, a minimal interaction determined by an extended derivative is an integral part of the local Landau theory and is symmetry-dependent. Representations with $\vec{k} \neq 0$ where vector $\vec{k}$ characterizes nontrivial translational properties of the order parameter (OP), are poorly described in field theory, whereas the subgroup of translations plays a significant role in physics. Studies of nontrivial representations of the subgroup of translations provide a basis for crystallography and physics of phase transitions. In the field theory, the compensating fields are determined by an abstract gauge group and do not depend on the representation of the wave function relative to the subgroup of translations. The trivial representations of the translation subgroup with $\vec{k} = 0$ are usually considered. Contrary to this, in the present theory the transformational properties of compensating fields are entirely determined by local transformational properties of the OP. We will show that for the Ginzburg-Landau model the OP local transformational properties are determined by a local transformation of the wave function under temporal translations $\hat{\tau}$:

$$\hat{\tau}\psi = e^{i\omega(x)\tau}\psi \ . \tag{1}$$

For the de Gennes' model, the OP local transformational properties are determined by its transformation at spatial translations $\hat{\vec{a}}$:

$$\hat{\vec{a}}\eta_l(\vec{X}) = e^{i\vec{k}_l(\vec{X})\vec{a}}\eta_l(\vec{X}) \ , \tag{2}$$

here $l$ gives the number of components of OP.

An assumption about locality of transformational properties of the OP (dependence $\vec{k}_l = \vec{k}_l(\vec{X})$) was made in [3], and gives a dependence of OP on the coordinates $\eta_l = \eta_l(\vec{X})$. Generally, in the inhomogeneous Landau theory, not only the value of OP can depend on coordinates, but the transformational properties of OP determined by the vector $\vec{k}$ also depend on coordinates as well. Such OP describes an inhomogeneous low-symmetry state where vector $\vec{k}$ is defined in each macroscopically small region marked by macrocoordinate $\vec{X}$. At this region, the local Landau potential $\Phi(\vec{X})$ can be constructed. Obviously dependence of the OP on $\vec{X}$ in the inhomogeneous models [1,2,4,5] implies that the local Landau potential does exist. Otherwise, a dependence of the OP on the coordinate would be impossible since the OP is equivalent to the coefficients in the Fourier $\vec{x}$ coordinate transformation of state density. Later, we shall describe inhomogeneity of the functions as a dependence on $\vec{X}$ coordinate (not $\vec{x}$ coordinate), to stress a macroscopic nature of the inhomogeneity.

In fact, the first attempt to study the model with the local transformational properties of OP in the Landau theory has been made by de Gennes [2]. He used a director $\vec{n}$ (unit normal to SmA layers) as an independent variable when describing the elastic properties of SmA. For a deformed state of SmA, the director $\vec{n}$ was assumed to be dependent on $\vec{X}: \vec{n} = \vec{n}(\vec{X})$. The OP's vector is equal $\vec{k} = 2\pi\vec{n}/d$, so $\vec{k}$ depends on $\vec{X}$. Here $d$ is the distance between the layers in SmA. Note that de Gennes model assumed $d = const$.

## 2. Compensating Field

For models with $\vec{k} \neq 0$, an inhomogeneity of the sort $\vec{k}_l = \vec{k}_l(\vec{X})$ results in the nontrivial transformations of the OP derivative under unit translations [3]:

$$\hat{\vec{a}}\frac{\partial \eta_l}{\partial X_j} = \frac{\partial}{\partial X_j}\left(e^{i\vec{k}^l(\vec{x})\vec{a}}\eta_l(\vec{X})\right) = e^{i\vec{k}^l(\vec{x})\vec{a}}\left[i\frac{\partial \vec{k}^l(\vec{X})\vec{a}}{\partial X_j}\eta_l(\vec{X}) + \frac{\partial \eta_l(\vec{X})}{\partial X_j}\right]. \quad (3)$$

As it is known [4], the inhomogeneous Landau potential is an invariant function constructed on the basis of OP and its spatial derivatives with respect to the symmetry of high-symmetry phase. It follows from (3) that the translation operator maps spatial derivatives of the OP, $\partial \eta_l / \partial X_j$, onto space of the OP itself, $\eta_l(\vec{X})$. By definition, components of the OP are the eigenfunctions of the translation operator (2). To construct invariants of the subgroup of translations which include spatial derivatives, we need first to construct the diagonal basis for the translation operator that includes the spatial derivatives of the OP. We apply the procedure of the derivative expansion which is proposed in the gauge field theory [6]. We extend spatial derivatives by introducing additional compensating fields into the derivative as such terms that the extended derivatives would be the eigenfunctions of the translation operator. According to (3) the compensating fields should be defined up to a gradient of the vector function. Let us show that this compensating field is a tensor.

The extended derivative is written down as:

$$D_j^l \eta_l = \left(\frac{\partial}{\partial X_j} - i\gamma \sum_p A_{pj}^l\right)\eta_l, \quad (4)$$

where action on $\gamma A^l_{pj}$ of the translation operator is determined in such a way that extended derivative (4) should be the eigenfunction of the translation operator:

$$\hat{\bar{a}}_q(\gamma A^l_{pj}) = \gamma A^l_{pj} + \delta_{pq} \frac{\partial k^l_p}{\partial X_j} a_q. \qquad (5)$$

Here, $A^l_{pj}$ is the compensating field, and $\gamma$ is a phenomenological charge. A dimension of the compensating field $A^l_{pj}$ is associated with the dimension of the vector $\vec{k}^l$ and is a second-rank tensor, since $A^l_{pj}$ should be transformed like $\partial k^l_p / \partial X_j$ with transformations from point symmetry group of reciprocal space (5). Vectors in the star $\{\vec{k}\}$ of the irreducible representation (IR) are dependent, since they are obtained from singl vector $\vec{k}$ by operations of the point symmetry group. Therefore, single tensor field $A_{pj}$ can be chosen such that it will compensate all vectors $\vec{k}^l$ of the IR. For example, for the six-beam basis of the icosahedron [3]:

$$\vec{k}_1 = (0,1,h),\ \vec{k}_2 = (0,\bar{1},h),\ \vec{k}_3 = (h,0,1),\ \vec{k}_4 = (h,0,\bar{1}),\ \vec{k}_5 = (1,h,0),\ \vec{k}_6 = (\bar{1},h,0),$$

where $h = (\sqrt{5}+1)/2$, the extended derivatives take the form:

$$D^1_j \eta_1 = \left[\frac{\partial}{\partial x_j} - i\gamma(A_{2j} + hA_{3j})\right]\eta_1, \qquad D^2_j \eta_2 = \left[\frac{\partial}{\partial x_j} - i\gamma(-A_{2j} + hA_{3j})\right]\eta_2,$$

$$D^3_j \eta_3 = \left[\frac{\partial}{\partial x_j} - i\gamma(hA_{1j} + A_{3j})\right]\eta_3, \qquad D^4_j \eta_4 = \left[\frac{\partial}{\partial x_j} - i\gamma(hA_{1j} - A_{3j})\right]\eta_4, \qquad (6)$$

$$D^5_j \eta_5 = \left[\frac{\partial}{\partial x_j} - i\gamma(A_{1j} + hA_{2j})\right]\eta_5, \qquad D^6_j \eta_6 = \left[\frac{\partial}{\partial x_j} - i\gamma(-A_{1j} + hA_{2j})\right]\eta_6.$$

The extended derivatives with opposite vector $\vec{k}$ take the form: $\tilde{D}^l_j \eta_l = (D^l_j \eta_l)^*$, here $^*$ means a complex conjugate. While constructing the compensating field in (6) we took into account that the change of coordinates of the vector $\vec{k}(\vec{X})$ in reciprocal space $\mu_i(\vec{X})$ is equivalent to the change of the values of basis vectors $\vec{k}(\vec{X}) = \mu_i \vec{b}_i(\vec{X})$ of reciprocal space. As a result, the extended derivatives $D^l_j \eta_l$ of the OP are the eigenfunctions of the translation operator $\hat{\bar{a}}$, and are transformed like the OP in (2). It is worth to note that the extended derivatives (6) contain tensor components but not the contraction in the form of vector, as it seemed to be if one takes into account a linear dimensionality of the derivative itself. The point is that tensor $A_{pj}$ transforms over the first component as a vector under the transformations from point group along with the components of OP, while it transforms over the second component along with the derivative.

We can construct hence the eigenfunctions containing spatial derivatives of the OP for the translation operator. However, we introduce here an additional tensor compensating field that represents additional independent degrees of freedom. Besides, the symmetry-dependent interaction between this field and OP is determined by expressions (4) and (6). By analogy with the field theory [6], we call this interaction - the minimal interaction. As it will be shown below, phonons are related to these new fields, and a phenomenological description of the electron-phonon interaction in the BCS model is related to the extended derivatives.

### 3. Observables. Stresses and Dislocations

States described by an OP with local translational properties $\vec{k} = \vec{k}(\vec{X})$, have the following interpretation: inhomogeneous deformation of crystal lattice, which is generally accompanied by discontinuities and occurrences of dislocations. Dislocations as linear incompatibilities of lattice occur at the boundaries of crystal regions which have different periods. For example, for a two-dimensional lattice, if one combines two lattices with different periods we get discontinuities or incompatibilities of lattice for the vector $\vec{k}$ changing along the perpendicular direction $\vec{k}(\vec{X}) = (k_1(X_1, X_2), k_2(X_1, X_2))$. In a three-dimensional model, dislocations with lines perpendicular to the plane of reference are related to these incompatibilities. Here, the dislocations appear not under the random deformation of lattice but when two regions with different periods are "glued" that respects to representations about existence of local translational symmetry in the macroscopically small regions [3]. Therefore, the equations of state for the model with $\vec{k} = \vec{k}(\vec{X})$ must include the equations of the continuous dislocation theory. Based on these concepts, we can define the observed values in our model.

Using equations (4), (5) we are able to construct a translationally invariant inhomogeneous Landau potential as a function of the OP and its extended derivatives. In this model, the introduced compensating tensor field $A_{pj}$ is an independent variable, and a variation of the potential with respect to it must be equal to zero. Similar to electrodynamics, we have to take into account the translational invariants of the compensating tensor field. Translational invariants $A_{pj}$ take the form of the antisymmetric spatial derivatives over the second index:

$$e_{jkn}\left(\partial A_{pn}/\partial X_k\right). \tag{7}$$

where $e_{jkn}$ is third-rank antisymmetric tensor. Thus, the local Landau potential which is invariant with respect to elementary translation takes the form:

$$\Phi(\vec{X}) = \Phi(\eta_l \eta_l^*, D_j^l \eta_l D_j^{l*} \eta_l^*, e_{jkn} \partial A_{pn}/\partial X_k). \tag{8}$$

Here an interaction between the OP and the compensating field is included into the extended derivatives.

The physical interpretation for $A_{pj} \equiv \Sigma_{pj}$ is associated with the tensor potential of the stress field introduced by Kröner [7], who identified and described the analogy between magnetostatics and the continuous theory of stationary dislocations. The definition of the stress tensor:

$$\sigma_{pj} = e_{jkn}\left(\partial A_{pn}/\partial X_k\right) \tag{9}$$

is a condition of equilibrium for the solid state in the stationary conditions. Equations of state are obtained from the variation of the local Landau potential with respect to the components of the compensating field $\delta\Phi/\delta A_{pj}$, and coincide with the definition of dislocation in the continuous theory of dislocations [8]:

$$\rho_{pj} = e_{jkn}(\partial w_{pn}/\partial X_k). \tag{10}$$

Since $\partial\Phi/\partial\sigma_{pj} = w_{pj}$ is the tensor of elastic distortion, and it follows here from, according to (8) and (9), that $\rho_{pj} = \partial\Phi/\partial A_{pj}$ is the density of dislocations (10). Here the Burgers vector is defined as an integral $b_i = \int_{S_L} \rho_{ij} ds_j$ over surface limited by the contour $L$.

One should pay attention that for the stationary case, two curl expressions (9), (10) occur in the model. They are two observables, where one of them can be considered as a definition, and the other as an equation of the state. Therefore, the interpretation of the compensating field in the model is generally ambiguous [9].

In the Kröner's analogy with magnetostatics, there are analogies between the current density and the dislocation density, the electromagnetic potential and the stress field potential, the magnetic induction and the stress tensor [7]. Alternative definition of the compensating field $A_{pj} \equiv w_{pj}$ as an elastic distortion tensor means that expression (10) is a definition of the observed values of compensating field and (9) is a definition of the equation of state, where $\partial\Phi/\partial\rho_{pj} = \Sigma_{pj}$. There are some heuristic arguments in favor of this definition:

1) The elastic distortion tensor in the equation (10), as well as a compensating field tensor (5), is defined up to a gradient of the vector function – deformation tensor. As one knows, dislocations density does not depend on deformation tensor [8].

2) The elastic distortion tensor $w_{pj}$ is a second-rank tensor, and not second-rank pseudo-tensor as the stress field potential $\Sigma_{pj}$ in the equation (9), thus, it is not necessary to introduce pseudo-scalar charge $\gamma$ into an extended derivative [9].

3) The equilibrium condition (9) matches the equations of state better than the definition of the dislocation density (10) [8].

The argument 3) becomes obvious in the framework of the dynamical model [10]. The static equilibrium condition $\partial\sigma_{ij}/\partial X_j = 0$ must transform into the Newton equations for the case of dynamics:

$$\partial\sigma_{ij}/\partial X_j = \partial p_i/\partial T , \qquad (11)$$

where $p_i$ is the momentum. So, expression (9) in the dynamical model cannot be the definition of the stress tensor. In contrast with the expression (10), the expression (9) should be defined additionally. If the elastic distortion tensor $w_{pj}$ is chosen as a variable, then the stress tensor is defined by the equations of state. The definition of the stress tensor as a value conjugate to the elastic distortion tensor (or deformation tensor) is sufficient for the stationary theory of elasticity. In the dynamical model, a definition of the stress tensor corresponds to its original definition as a tensor whose divergence is force: $\partial\sigma_{ij}/\partial X_j = f_i$. If temporal derivatives of OP are considered in the theory [10] and are extended by the analogy with electrodynamics by introducing an additional vector potential field $\varphi_i$ compensating possible variations $\partial\vec{k}/\partial T$, then the equations of continuity of this model are the Newton equations (11). It is easy to demonstrate following the analogy with electrodynamics. In the electrodynamics, the equations of continuity take the form $\partial j_i/\partial X_i = \partial q/\partial T$ where $q$ is the charge density. Assuming that stress tensor corresponds to the current density, the momentum $p_i$ plays the role of a charge and a quantity conjugate to the vector potential $\varphi_i$. In electrodynamics, the law of conservation of charge is associated with the gauge group; in our model, the law of conservation of momentum is associated with the subgroup of unit translations $\hat{\vec{a}}_q$. The equations of continuity with the compensating tensor field, written as the second Newton law (11), are the differential form of the law of conservation of momentum. So, we assume $A_{pj} \equiv w_{pj}$ and study the model where the current density is associated with the stress tensor, and the dislocation density is associated with the magnetic induction. Alternative choice of observables in the previous articles [3] is possibly justified for some special problems [7, 9], and most likely is due to Kröner's authority.

The most known model in the phenomenological Landau theory which uses an extended derivative and takes into account the elastic dislocations at deformed state is the de Gennes model. de Gennes made an attempt to construct a phenomenological potential for the SmA similar to the Ginzburg-Landau potential in order to describe the effect of stress field screening by elastic dislocations, similar to the Meissner effect [1]. The de Gennes phenomenological potential has the form

$$\Phi = \alpha|\psi|^2 + \frac{\beta}{2}|\psi|^4 + \left|(\vec{\nabla} - ik_0\delta\vec{n})\psi\right|^2 + \Phi(\vec{n}) \tag{12}$$

where $\Phi(\vec{n}) = c_1(\vec{\nabla}\cdot\vec{n})^2 + c_2(\vec{n}\cdot\vec{\nabla}\times\vec{n})^2 + c_3(\vec{n}\times(\vec{\nabla}\times\vec{n}))^2$ is Franck potential.

However, he selected a compensating field in the extended derivative to be a vector based on dimensionality of the derivative itself, and did not check the translational invariance of the constructed smectic potential (12). It is easily seen that the de Gennes potential is not invariant with respect to the unit translation operator (2, 3) since the vector field cannot compensate variations of the director $\vec{n}(\vec{X})$ in three dimensions. As it is known, in the de Gennes model (12), the problem of screening of the stress field cannot be solved because the Franck potential $\Phi = \Phi(\vec{n}(\vec{X}))$ build for a nematic was used [2, 12]. The point is that the Franck potential contains noncurless terms of the type of divergence of the director, which cannot lead to the vortex equations similar to Londons' equations [1]. It is probably impossible to describe the elastic properties of SmA by the Franck potential because it does not take into account the periodic structure of smectic [8]. Due to the fact that variation of the director leads to the dislocations, de Gennes suggested it for a phenomenological description of the deformed SmA. In fact, one can see that the integral along a closed contour $L$ $\nu_L = \oint_L \frac{\vec{n}}{d}d\vec{r}$ defines a number of dislocations $\nu_L$ enclosed by the contour. Here $d = 2\pi/k_0$ is a distance between the layers, $\vec{k}_0$ is a wave vector of a non-deformed SmA [2].

It has been shown above how to construct a phenomenological model with dislocation density in the Landau theory with the local transformation properties of OP $\vec{k} = \vec{k}(\vec{X})$. The present model that includes a compensating tensor field to describe a deformed SmA, is free of deficiencies of the de Gennes model [13]. A vortex structure of the tensor equations of state for the potential (8), results in equations similar to London equations [14] and to a description of the screening effect of the stress field by elastic dislocations. We should also note that appearance of the screening effect does not depend on the choice of interpretation of a compensating field as a stress field potential or as an elastic distortion.

4. **Ginzburg-Landau Model**

By analogy with the spatial translations let us consider the model with local transformational properties of the OP at temporal translations:

$$\hat{\tau}\psi = e^{i\omega(x)\tau}\psi. \qquad (1)$$

In this case the extended derivative has the form

$$D_j\psi = \left(\frac{\partial}{\partial X_j} - ieA_j\right)\psi, \qquad (13)$$

where $e$ is phenomenological charge, and the compensating field $A_j$ is transformed under the operator $\hat{\tau}$ as:

$$\hat{\tau}(eA_j) = eA_j + \frac{\partial\omega}{\partial X_j}\tau. \qquad (14)$$

According to (14), the field $A_j$ is transformed as $\partial\omega/\partial X_j$ and, therefore, changes its sign upon inversion of time:

$$\hat{I}_\tau(A_j) = -A_j, \qquad \hat{I}_\tau(\psi) = \psi^*. \qquad (15)$$

We can construct so a local phenomenological Landau potential which is invariant with respect to elementary temporal translations:

$$\Phi = a|\psi|^2 + \frac{b}{2}|\psi|^4 + \lambda\left|(\vec{\nabla} - ie\vec{A})\psi\right|^2 + \Phi_e(\vec{\nabla}\times\vec{A}). \qquad (16)$$

Expression (16) is similar to the Ginzburg-Landau potential up to coefficients:

$$\Phi = a|\psi|^2 + \frac{b}{2}|\psi|^4 + \frac{\hbar^2}{4m}\left|(\vec{\nabla} - \frac{2ie_{el}}{\hbar c}\vec{A})\psi\right|^2 + \frac{1}{8\pi}\vec{B}^2, \qquad (17)$$

where $\vec{\nabla}\times\vec{A} = \vec{B}$ is a magnetic field. The phenomenological charge $e$ is related to the physical constants and the electron charge as $e = e_{el}/\hbar c$. If the choice of the coefficients in front of derivative in the expression (17) is conditional [14], then the charge doubling in the derivative of the Ginzburg-Landau model [15] is necessary to get a correct expression for the current density in the magnetic field. The current density for the potential (17) is given by [14]:

$$\vec{j} = -\frac{ie_{el}\hbar}{2m}(\psi^*\nabla\psi - \psi\nabla\psi^*) - \frac{2e_{el}^2}{mc}|\psi|^2\vec{A}. \qquad (18)$$

Then the Londons equations (with the assumption $|\psi|^2 = const$) have the form:

$$\vec{\nabla}\times\vec{j} = -\frac{e_{el}^2 n_s}{mc}\vec{B}. \qquad (19)$$

Here $n_s = 2|\psi|^2$ is the density of superconducting electrons.

A doubling of charge in the derivative (17) is enforced neither by the phenomenology [1, 14] no by the symmetry requirements in the Landau theory, but it is a consequence of the BCS theory. We show below that a correct expression for the current density in (18) can be obtained if one chooses a correct representation of the OP with the local transformation properties relative to both temporal (1) and spatial (2) translations, and the charge doubling in extended derivative is not necessary.

It is known that the extended derivative in the Ginzburg-Landau potential (17) was adapted from the field theory (electrodynamics). On the other hand, we get similar expression (16) when constructing a local Landau theory for (1). The question arises how the minimal interactions in the field theory and the Landau theory are related. In the Abelian gauge model of the field theory [6, 11] with the gauge group U(1) that corresponds to electrodynamics, the Lagrangian should be invariant under the local gauge transformation of phase of the wave function $\hat{g}\psi = e^{i\alpha(X)}\psi$. The compensating field of such a model is transformed under the local gauge group as

$$g(eA_j) = eA_j + \partial\alpha/\partial X_j, \qquad (20)$$

and extended derivative has the same form as (13). In the Abelian gauge field theory change of the sign of the compensating field in the extended derivative under the time inversion is postulated. This is due to the fact that a scalar phase factor $\alpha(\vec{X})$ is invariant under temporal inversions, being the parameter of the gauge group of electrodynamics U(1). Therefore the transformation properties of the compensating field upon the temporal inversion in the U(1) model [11] have to be additionally defined. On the other hand, knowing that the electromagnetic vector potential $A_j$ changes its sign upon inversion of time, one can state that such a representation of OP, with the electromagnetic potential $A_j$ acting as a compensating field, must have the local transformational properties (1) at temporary translations. Indeed, the extended derivative (13) has the terms that are transformed in the different ways under the inversion of time $\left(\partial/\partial X_j - ieA_j\right)$, and the explanation must be given to this. Additional transformational properties of the compensating field in the extended derivative are associated with the transformational properties of the OP itself (1), (14). The compensating field in the Ginzburg-Landau model is transformed together with the OP upon the inversion of time $\hat{I}_\tau(\psi) = \psi^*$, being in agreement with the representation (1).

Similar to this, if the transformational properties of the OP at spatial translations are local, then the extended derivatives (4), (6) are not vectors. A compensating field occurs in the extended derivative (4), (6) not as a contraction of a tensor by the first index, but as a tensor with the first index which transforms together with the OP vector $\vec{k}$ and with the second index which transforms together with spatial derivatives. Therefore, in the present model, the extended derivative is not a vector by definition. We conclude that a dimensionality of the extended derivative depends on a choice of the local IR. Thus model with the local group U(1) [6, 11] is not equivalent to the models (1) and (2). The local representations of the translation subgroup are not equivalent to the local abstract gauge groups because the dimensionality of the compensating fields at transformations in the reciprocal space is determined by their transformations in (5), (14), which is absent in (20). Hence, the non-trivial representations of the translational subgroup (1), (2) cannot be reduced to the gauge groups. The local representations of the temporal translations subgroup given here seem to be more suitable for the description of electrodynamics on the framework of the group theory (U(1) together with inversion of time $\hat{I}_\tau$). By definition, the transformations in the inner abstract space for non-Abelian groups are not related to the space-time transformations. Therefore using the non-Abelian groups to describe the states with the local translation symmetry in the Landau theory will not likely result in equivalent models.

The models with local gauge groups for the fields with $\vec{k}=0$ and the models with local representations of translations subgroup with $\vec{k}=\vec{k}(\vec{X})$ describe different objects. In the former case the extended derivative is a vector, while in the latter it contains tensor compensating fields.

## 5. Superconductivity and Electron-Phonon Interaction

As it is known, the high-temperature superconducting (HTS) states are inhomogeneous and anisotropic (for reviews see, e.g., [16, 17] and more references there). However it may be assumed that in the HTS states, a short-range crystallographic order exists, hence, a local Landau potential may be constructed. I.e., we have to construct the Ginzburg-Landau potential for the states with $\vec{k}=\vec{k}(\vec{X})$. So, representation of the OP for HTS state must have local transformational properties both (1) and (2). Since the dimensionality of the extended derivative depends on the chosen representation of translations subgroup, the minimal interaction with one extended derivative that contains both fields compensating variations of vector $\vec{k}(\vec{X})$ (2), and fields compensating local properties of OP at temporal translations (1) can be constructed.

In this case, the extended derivative will include linear terms of both elastic distortion tensor $A_{pj}$ and electromagnetic vector-potential $A_j$. Then, for the IR with $\vec{k} \neq 0$ and $\psi_{\vec{k}_l} = \psi_{\vec{k}_l(\vec{X})}$, we assume that

$$\hat{\tau}\psi_{\vec{k}_l} = e^{i\omega(x)\tau}\psi_{\vec{k}_l}, \; \hat{\tau}\psi^*_{\vec{k}_l} = e^{-i\omega(x)\tau}\psi^*_{\vec{k}_l}, \qquad (21)$$

and the inversion of time acts as $\hat{I}_\tau(\psi_{\vec{k}_l}) = \psi^*_{\vec{k}_l}$ for each $\vec{k}_l$ in $\{\vec{k}_l\}$. Taking into account that nontrivial transformations of the OP at temporary translations doubles the dimensionality of the IR $\{\psi_{\vec{k}_l}, \psi^*_{\vec{k}_l}\}$, so in this case, the quadratic invariants of the OP take the form:

$$I = \sum_l \psi_{\vec{k}_l}\psi^*_{-\vec{k}_l} + \psi^*_{\vec{k}_l}\psi_{-\vec{k}_l}. \qquad (22)$$

According to (4), (12), the extended derivatives read:

$$D_j\psi_{\vec{k}_l} = \left(\frac{\partial}{\partial X_j} - i\gamma\sum_p A^l_{pj} - ieA_j\right)\psi_{\vec{k}_l},\; D_j\psi^*_{\vec{k}_l} = \left(\frac{\partial}{\partial X_j} - i\gamma\sum_p A^l_{pj} + ieA_j\right)\psi^*_{\vec{k}_l}, \qquad (23)$$

$$D_j\psi_{-\vec{k}_l} = \left(\frac{\partial}{\partial X_j} + i\gamma\sum_p A^l_{pj} - ieA_j\right)\psi_{-\vec{k}_l},\; D_j\psi^*_{-\vec{k}_l} = \left(\frac{\partial}{\partial X_j} + i\gamma\sum_p A^l_{pj} + ieA_j\right)\psi^*_{-\vec{k}_l}.$$

Hence, quadratic gradient invariants, which are responsible for minimal interaction, may be written in the form:

$$\Phi_i = \lambda\sum_{jl} D_j\psi_{\vec{k}_l}D_j\psi^*_{-\vec{k}_l} + D_j\psi^*_{\vec{k}_l}D_j\psi_{-\vec{k}_l}. \qquad (24)$$

Here $\lambda$ is a scalar. Expressions for $I$ and $\Phi_i$ (22), (24) do not take into account the anisotropy because the latter does not change the conclusions below. In fact, the extended derivative (23) contains components of the 4-tensor, with the first three of them compensating for inhomogeneous changes of the OP at spatial translations, while the fourth one does this for temporary translations.

For such model, the local Landau potential has the form

$$\Phi = \Phi_L + \Phi_i + \Phi_e(\vec{\nabla} \times \vec{A}) + \Phi_f(e_{ijk} \partial A_{pj}/\partial X_k), \qquad (25)$$

where $\Phi_L = aI + b/2\, I^2$ is the conventional Landau potential depending only on the OP components, $\Phi_i$ is a part of the potential containing the OP derivatives and is responsible for the minimal interaction with the compensating field. $\Phi_e(\vec{\nabla} \times \vec{A})$ and $\Phi_f(e_{ijk} \partial A_{pj}/\partial X_k)$ are quadratic forms corresponding to the electromagnetic free energy and the elastic one, resp. The Maxwell equations entering the equations of state have the form:

$$\delta\Phi/\delta\vec{A} = \partial\Phi_i/\partial\vec{A} - \vec{\nabla} \times (\partial\Phi_e/\partial(\vec{\nabla} \times \vec{A})) = 0. \qquad (26)$$

The Londons equations in the model (21) arise from (23, 24). In fact, because $A_j$ occurs only in $\Phi_i$ (25), then the vector of current contains only terms that are linear in $A_j$ with coefficients $\psi_{\vec{k}_l}\psi^*_{-\vec{k}_l} + \psi^*_{\vec{k}_l}\psi_{-\vec{k}_l} = 2|\psi|^2$. Moreover, the vector of current does not contain $A_{pj}$ because the components $A_{pj}$ occur in the equation for the current with zero coefficients: $\psi_{\vec{k}_l}\psi^*_{-\vec{k}_l} - \psi^*_{\vec{k}_l}\psi_{-\vec{k}_l} = |\psi|^2 - |\psi|^2 = 0$, in accordance with (23, 24).

So, in the present model (21, 22) the elastic fields $A_{pj}$ do not enter the Maxwell equations explicitly due to symmetry of expressions (23, 24). The said above means that within the approximation $|\psi|^2 = const$ the Londons equations are correct for the model (21) [14]. Moreover, for the effective potential depending only on the four OP components $(\psi_{\vec{k}}, \psi_{-\vec{k}}, \psi^*_{\vec{k}}, \psi^*_{-\vec{k}})$ with the designated value of vector $\vec{k}$, doubling of the coefficient before electromagnetic potential in the expression for the current takes place, that corresponds to the idea about electron pairing in a superconductive state. Obviously, this effective potential describes a couple of electrons with equal antiparallel momentum. We should note that equations of state always have solutions when all the OP components except the components with designated opposite vectors $\vec{k}$ are equal to zero [13].

The fact that the vector of current density does not depend on the tensor of elastic distortion means that there is no dissipation in this model. It is easy to see that since the current does not depend on the tensor of elastic distortion, so the inner stresses do not depend on electromagnetic potential in the present model.

Let us analyze potential (25) when the OP is equal to zero. Equations of state for the free field $A_{pj}$ are $\delta\Phi_f/\delta A_{pj} = 0$. For the dynamic case $\Phi_f = \Phi_f(e_{ijk} \partial A_{pj}/\partial X_k, -\partial\phi_p/\partial X_i - \partial A_{pi}/\partial T)$ because $-\partial\phi_p/\partial X_i - \partial A_{pi}/\partial T$ is an invariant of the translation group under the condition $\widehat{\bar{a}}_q(\gamma\phi_p) = \gamma\phi_p - \delta_{qp}\dfrac{\partial k_p}{\partial T}a_q$ [10] and equation (5), where $\phi_p$ is a potential conjugate to momentum. These equations of state are equivalent to the wave equations for the free electromagnetic field $\delta\Phi_e/\delta A_j = 0$, where

$\Phi_e = \Phi_e(\vec{\nabla} \times \vec{A}, -\vec{\nabla} \cdot \phi - \partial \vec{A}/\partial T)$. Here, the exact wave functions are solutions of equations of state for the free field $A_{pj}$ (the necessary conditions to obtain wave solution, the so-called gauge conditions, will be discussed below). As it is clear from what follows, the tensor $A_{pj}$ corresponds to the phonon potential and is responsible for the phenomenological description of the electron-phonon interaction in (25). Following analogy with the electrodynamics, one could tell that inner stresses $\partial \Phi/\partial A_{pj} = \sigma_{pj}$ are the phonon sources. The electron-phonon interaction in (23), (24) has the same structure as the electromagnetic interaction: it is minimal and determined by symmetry.

We should note that in the local Landau theory other minimal interactions do not exist, because we took into account the locality of the OP transformational properties with respect to the both temporal and spatial translations. The last conclusion results from the mechanism of the IR construction in the Landau theory itself. First, we expand the function into Fourier series and hold the fixed OP transformation properties with respect to the translation subgroup, and then we choose the OP relatively to the point subgroup. Thus, even if subgroup of rotations is infinite (for example, an infinite subgroup of rotations around the main optic axis exists for the deformed SmA), then it caused not the additional continuous local IR parameter characterizing rotation group, but the infinite-dimensional OP. As was shown above, it is the locality of the continuous IR parameters $\vec{k}(\vec{X})$, $\omega(\vec{X})$ of the spatial group that results in the minimal interaction in the Landau theory.

This the phenomenological model (21) agrees with the BCS model and describes not only the electromagnetic, but also the electron-phonon interaction. The latter being taken into account results in the electron pairing and superconducting state, where the current does not depend on the inner stresses. The extension of derivative associated with the introduction of the phonon tensor potential $A_{pj}$ in (23) is caused by the local translational symmetry and anisotropic of the HTS samples.

For conventional superconductors, the electron-phonon interaction should also be present in the phenomenological description of inhomogeneous states. As we noted before, the Ginzburg-Landau functional describes the inhomogeneous states and implies local homogeneity. Thus, the functional must take into account the locality of the OP transformational properties with respect to elementary translations, which results in the electron-phonon interaction.

## 6. Overview and hypothesis about the sound.

In the Ginzburg-Landau phenomenological theory of superconductivity [1] the electron-phonon interaction was not taken into account explicitly and the phonon potential did not exist in its context. In accordance with the BCS theory, it is the electron-phonon interaction that is responsible for the electron pairing and phenomenon of superconductivity. Taking into account conventional interaction between the superconducting OP and the lattice oscillations as interaction with elastic degrees of freedom (for example, with displacements written as deformation tensor or with stresses tensor) would result in violation of the Londons equations and, consequently, the Meissner effect in this model. As shown in the present work, choice of the IR model with $\vec{k} = \vec{k}(\vec{X})$ generates a non-trivial interaction between the superconducting OP and the phonon potential, similar to electromagnetic interactions, and description of superconductivity in the Landau phenomenological theory.

Let us note that to get the correct expressions for the current density and the Londons equations in the present phenomenological theory we should not resort to the charge doubling by hands in the extended derivative and the wave function renormalization as was done in [14, 15].

An appropriate choice of the representation with $\vec{k} \neq 0$ (21) solves this problem in the local Landau theory.

The present results are related to the construction of minimal interaction in the local Landau theory. The idea about extension of derivatives for construction of translation invariants of the Landau inhomogeneous potential is borrowed from the gauge field theory [6, 11]. Let us do a dimensional analysis of the compensating fields. We can do it in the local Landau theory (5, 14), contrary to the field theory (20) [11]. Whatever are reasons for occurrence of the local symmetry in the low-symmetry state, one could assume (in accordance with the local homogeneity [10]) that the group itself is local but not the parameters of representation, i.e., $\vec{a} = \vec{a}(\vec{X})$ in (5) and $\tau = \tau(\vec{X})$ in (14), instead of $\vec{k} = \vec{k}(\vec{X})$ and $\omega = \omega(\vec{X})$, respectively. Then, obviously, the elastic distortion tensor $w_{ij}$ is defined up to deformation $\partial a_i / \partial X_j$. Also, the potential vector field $\phi_i$ introduced in [10] is conjugate of momentum $p_i$ and is defined up to speed $\partial a_i / \partial T$. Similar to this, the electromagnetic potential it defined up to $\partial \tau / \partial X_j$ that gives evidence of its relativistic origin. In the constructed model, we have the next conjugations: $\partial \Phi / \partial w_{ij} = \sigma_{ij}$, $\partial \Phi / \partial \phi_i = p_i$. Alternatively, as it is known, the stress tensor is a conjugate to deformations tensor and momentum is a conjugate to velocity vector. Hence, varied compensating fields have the same mathematical and physical dimensionality as the deformation tensor and the velocity vector. However, they are defined as independent fields but cannot be represented by spatial and temporal derivatives that are deformation tensor and speed vector. Moreover, the combinations vortices of compensating fields defined up to a deformation tensor and a vector of velocity due to symmetry requirements, are the observables in present model. In fact, the elastic distortion tensor appears in the continuous dislocations theory as a tensor with nonzero curl for states where displacements are not defined in general. That is why the dislocation density does not depend on the deformation tensor. I.e., we consider as a variable the tensor field which cannot be presented as a derivative of vector function and we shall examine it up to a vector function derivative. In the classic elastic theory where the displacement vector $u_i$ occurs as a basic variable, the deformation tensor is an invariant of the global translation subgroup. It is true in the continuous media approximation. However, for a crystal in the deformed state, a general situation with the local translation subgroup $\vec{a} = \vec{a}(\vec{X})$ is realized. Since by definition of displacement vector $\hat{a}_i(u_i) = u_i + a_i(\vec{X}) - a_{i0}$ (where $a_{i0}$ is a period in the non-deformed state), then $\hat{a}_i(\partial u_i / \partial X_j) = \partial u_i / \partial X_j + \partial a_i / \partial X_j$ and, therefore, the deformation tensor is not an invariant of the elementary translation subgroup. Therefore, in this case, we should describe elastic properties of crystal via elastic distortion tensor that plays a role of compensating field (4), (5). The electrodynamics provides example of what this types of models could result in. To our opinion, a study of the dynamic models with compensating tensor fields should yield a phenomenological description of sound. The BCS model, as well as the phenomenon of superconductivity itself, is the important arguments to suppose this hypothesis.

There are some heuristic arguments based on the electromagnetic analogue in favor of the phenomenological sound description via the phonon potential. As it is known, an electromagnetic wave consists of two components: magnetic and electric fields, with the electromagnetic potential satisfying the d'Alembert wave equation. Similar to the electromagnetic ($A_j, \phi$) potential and the phonon ($A_{pj}, \phi_i$) potential, the sound would occur as a pair: the antisymmertric combination of $e_{ijk} \partial A_{pk} / \partial X_j$ and the gradient $-\partial \phi_i / \partial X_j - \partial A_{ij} / \partial T$, that change in accordance with wave law (here $\phi_i$ is field of velocity, and a quantity conjugate to momentum; it is independent variable, in analogy to the distortion tensor).

Empirically, the gradient of the field of velocity can be easily observed in any musical instrument, for example, in the oscillating string with fixed ends. However, to our opinion, the most interesting are wind instruments that all have widening or holes at their end. The gradient of the field of velocity clearly comes about from air movement from wind instruments as a sound. The dislocation density also occurs when the sound is generated. For example, to generate a sound from a string, we stretch it normally to the tension force, but not along it. It is related to the mechanism of dislocation production in the inhomogeneous state that was described in the beginning of section 3. A force defined as a stress divergence has the same components as the Burgers vector which characterizes the lattice incompatibilities (9, 10). Therefore, to create a dislocation, we have to produce non-homogeneity in the direction perpendicular to the applied force. Also, the sound always appears along with the formation of linear defects.

There is a natural question: how the phonon potential and its invariants characterizing the phonon field intensity are related to the mechanical oscillations associated with the sound [8, 18]? To our opinion, the answer comprises an additional condition for the compensating fields. As is known, the wave solutions of Maxwell equations for a free electromagnetic field exist with an additional Lorentz gauge condition. For the three-parameter group with $\vec{k} = \vec{k}(\vec{X})$, the additional conditions that should be applied to the compensating fields to obtain wave solutions for equations of state ($A_{pj}, \phi_i$), have the following form:

$$\frac{\partial A_{pj}}{\partial X_j} + \frac{1}{c^2}\frac{\partial \varphi_p}{\partial T} = 0 \qquad (29)$$

where $c$ is a wave velocity (for simplicity, we do not take into account anisotropy). Contrary to the Lorentz condition, the squared sound velocity is present in (29). It is related to relativistic normalization of the electric potential $g(e\phi) = e\phi - \frac{1}{c_e}\partial\alpha/\partial T$ under the gauge transformation (20), where $c_e$ is speed of light. It is easy to see that the velocity $c$ is a function of phenomenological coefficients of the potential $\Phi_f = \Phi_f(e_{ijk}\partial A_{pj}/\partial X_k, -\partial\phi_p/\partial X_i - \partial A_{pi}/\partial T)$. In fact, $c^2 = a_p/a_\rho$, where $a_p$ is a coefficient before quadratic invariant composed from spatial derivatives of phonon potential $e_{jkn}(\partial A_{pn}/\partial X_k)$, and $a_\rho$ is coefficient in front of the quadratic invariant composed from the gradient $-\partial\phi_p/\partial X_j - \partial A_{pj}/\partial T$. Phenomenological coefficients $a_p$ and $a_\rho$ have the physical dimensionality as pressure and density, respectively. Similar to field theory [11], the condition (29) caused d'Alembert equation for the group parameter $\vec{k}$:

$$\Delta k_p - \frac{1}{c^2}\frac{\partial^2 k_p}{\partial T^2} = 0 \qquad (30)$$

(it follows from the requirement of translation of invariance of (29)). This means that we study a model where the components of vector $k_p$ evolve by the wave law with the speed $c$. The innverse is also true, if vector $\vec{k}$ satisfies the equation (30), then the field components ($A_{pj}, \phi_i$) are related by the additional condition (29). The field components satisfy the d'Alamber wave equation with the operator $\Delta - \frac{1}{c^2}\frac{\partial^2}{\partial T^2}$ where the phenomenological parameter $c$ is equal to parameter in equation (30). Varying of vector $\vec{k}(\vec{X})$ according to the equation (30) can be interpreted as varying of lattice period $\vec{a}(\vec{X})$ by the wave law. Hence, wave equations for free phonon field exist in the state where mechanical wave oscillations are realized (acoustic $c = const$). But the phonon field itself is not a mechanical media oscillations, and as it was shown before (23), the electron-phonon interactions cannot be described via displacements or the deformation tensor. Such mechanical description of the phonon potential would be similar to attempts of mechanistic description of the compensating field as the variation of a director in the de Gennes' model (12); such approach would be beneath criticism. All mentioned above gives us grounds to represent a sound as waves of translation invariant combinations of compensating fields. The sound is a wave existing under condition of mechanical oscillation in (30), similar to the electromagnetic wave.

## 7. Conclusion. Analogy with electromagnetism.

In conclusion, we shall write expressions for the three-parametric model with $\vec{k} = \vec{k}(\vec{X})$, similar to the expressions for Coulomb force and Lorentz force in one-parametric electrodynamics model, and then we shall study their relation to the Euler equations.
The expression
$$f_j = -p_i \partial \phi_i / \partial X_j \tag{31}$$
is analogue to the Coulomb force $qE_j = -q \partial \varphi / \partial X_j$. As it is known the same expression holds for the force in the context of the potential motion of liquid [18]. Substituting the force in the usual form $-p_i \partial v_i / \partial X_j$, where $\vec{v} \equiv \vec{\phi}$, to equations of motion, and taking also into account that for the isotropic media $\sigma_{ij} = -\delta_{ij} p$ (where $p$ is a pressure), we obtain the Euler equations for the potential motion of liquid:
$$\rho \partial v_j / \partial T = -p_i \partial v_i / \partial X_j + \partial \sigma_{ji} / \partial X_i \tag{32}$$
or, in a more usual form [18], $\partial v_j / \partial T = -v_i \partial v_i / \partial X_j - \frac{1}{\rho} \partial p / \partial X_j$. In stationary case this equation yields the Bernoulli equations. It follows from (32) that gradient of the velocities field is intensity of momentum, in agreement with the above hypothesis about the sound.

The important parallel with electrodynamics is traced for the Lorentz force $f_i = e_{ijk} j_j B_k$, where $B_k = e_{klm} \partial A_m / \partial X_l$. In the expression for the Lorentz force we substitute current density by the stress tensor and magnetic field by the dislocations density:
$$f_i = e_{ijk} \sigma_{nj} \rho_{nk}, \tag{33}$$
here $\rho_{nk} = e_{klm} \partial A_{nm} / \partial X_l$. We obtain the force applied to dislocations when their density is distributed continuously. Similar expression holds in the theory of elasticity [8, 19]:
$$f_i = e_{ikj} \tau_k b_n \sigma^e_{nj}. \tag{34}$$
It is the known Peach-Koehler force (here summation of stresses tensor and antisymmetric tensor is made by the second index [19], but not by the first one like in [8], due to the definition of the force as the divergence of the stress tensor by the second index). Taking into account that the tensor $e_{ikj}$ is antisymmetric, the permutation of neighboring indexes causes sign change, reflecting transition from inner stresses in (33) to outer stresses in (34). One should note that the form of the expression $\rho[\vec{v} \times [\vec{\nabla} \times \vec{v}]]$ describing the vortex flow in the Euler equation, is implicitly present in the expression (33) under condition (29). In fact, it follows from the Lorentz condition (29) (that has the form of the continuity equation), that the distortion tensor $A_{nm}$ can be represented as a current of field of velocity $-c^{-2}\phi_n v_m$, where $v_m$ is a velocity of current. Substituting $\sigma_{ij} = -\delta_{ij} p$ into the force expression (33) and assuming $\vec{v} = \chi\vec{\phi}$ for liquid media, one obtains the expression $\chi p c^{-2}[\vec{v} \times [\vec{\nabla} \times \vec{v}]]$, where $\vec{v} \equiv \vec{\phi}$. It is equivalent to the vortex term in the Euler equations when $c^2 = \chi p/\rho$. This is the known Newton equation for velocity of sound. We shall note that the Peach-Koehler force is exactly equal to the curl term in the Euler equations if $v_j \partial \phi_j/\partial X_i - v_i \partial \phi_j/\partial X_j = 0$. This condition is valid for sound solutions studied in [18]. The above considerations amount to the derivation of the sound velocity. The latter appears in the theory as a parameter in the additional conditions (29) which are necessary to obtain the wave solution.

In a general case for derivation of the formula for the sound velocity the Hooke law and the condition (29) suffice. Indeed the stress tensor is proportional to the distortion tensor $\sigma_{ij} = -K A_{ij}$, with is a generalization of the deformation tensor. According to the continuity equation (11) $\sigma_{ij} = p_i v_j = \rho v_i v_j$ and is $A_{ij} = -c^{-2}\chi v_i v_j$, as follows from (29) for isotropy condition. Thus we obtain the known expression for the sound velocity in gases and liquids $c^2 = \chi K/\rho$. It is obtained due to continuity equation (29), but the latter is not a conservation law.

We shall write down the Euler equations as:
$$\partial(\rho v_j)/\partial T = -\partial(p_i v_j)/\partial X_j + \partial \sigma_{ji}/\partial X_i . \qquad (35)$$
It follows from (35) that the Euler equations are the superposition of two forces that are different in general. In fact, each term in the right part of (35) is the force definition. That is why in the model with $\vec{k} \neq 0$, the equations of motion can be obtained as superposition of the Newton continuity equations (11) and the forces similar to the Coulomb and Lorentz forces (31), (33) that are related with the homogeneity of the macro-space $\{\vec{X}\}$. It became possible due to coincidence of the physical dimensions of continuity equation and differential form of the momentum conservation law in this model. In the electrodynamics, these dimensions are different. To our opinion, the surprising thing is that in the Euler equations the force obtained by the differentiation in partial derivatives can be also obtained on the basis of mathematical model with extended derivatives. This force is equivalent to the force acting to the particle in the phonon field under condition (29).

Similar to the Maxwell equations, equations of state of phonon field contain linear combinations of the compensating fields. Such equations have periodical wave solutions when the OP is absent, or when the source is periodical wave function. Thus, the phonon fields describing the sound, have periodical wave solutions. Then, to obtain the exact wave solution of equations of state, we do not neglect the non-linear term in the Euler equations [18]. It became possible due to accounting for additional generalized degrees of freedom in the form of compensating (phonon) fields. Approximate wave solutions of the non-linear Euler equations are not valid for the description of the sound. In the case of approximate wave solutions, the sound frequency depends on its amplitude, and the sound tone would change while power increases, that is not observed. The Euler equations describe not a sound, but variations of density and velocity of continuous medium generated by the sound oscillations. To our opinion, the sound waves should be described by the wave equations for the tensor compensating field, similar to the electromagnetic waves.

Concerning the analogy between the local Landau theory and general theory of relativity (GTR), we note that non-homogeneous lattice deformations can be described by a local metric in macroscopically small space characterized by macro-coordinate $\vec{X}$. However, when differentiating in the space $\{\vec{X}\}$, there are no reasons to introduce a local metric, because the symmetry group $\{\vec{X}\}$ is global. In general, the occurrence of crystal lattice in the space $\{\vec{X}\}$ under deformation results in the local translational symmetry. In the Landay theory the representation, but not the symmetry group of high-symmetry phase, is local. However, when constructing the extended derivatives (for example, for model (6)), one can formally turn the change of coordinates of vector $\vec{k}$ into the change of value of basis vectors of reciprocal space. Here it is not relevant whether one should describe the local lattice symmetry via a metric tensor or via values of reciprocal space basic vectors. It is important that the constructed Landau potential should be translational invariant. The crystal lattice metric locality results in the local group of the Landau potential, but not in the local metric in macro-space $\{\vec{X}\}$ and the GTR model. In fact, the translational symmetry in the local space is the function of coordinates $\vec{X}$ of the macro-space that corresponds to the principle of local homogeneity [3]. Here, elements in $\{\vec{X}\}$ are invariant under the elementary translations of local space:

$$\hat{\vec{a}}_i \vec{X} = \vec{X}. \tag{36}$$

The last statement is mathematical expression of the principle of local homogeneity [10]. For the model (2), the transformation of coordinates (36) agrees with Noether theorem and conservation law (11). The continuity equations (11) can be easily obtained from the simultaneous equations of state, similar to continuity equations for current in the electrodynamics. Thus, the expression (35) definitely follows from Noether theorem and means that macro-coordinates $\vec{X}$ are independent under the lattice local deformations. In this model, deformations are described by compensating field $A_{pj}$. When the lattice is absent or when it can be neglected, the deformations in $\{\vec{X}\}$ are described in the context of the classical theory of elasticity of continuous medium [8].

The relationship between description of linear defects and GTR is analyzed in the review [20]. It has been suggested that states with dislocations are described by local reper $e_j^p$ that is independent tensor function being the generalization of spatial derivatives of vector field of displacements $\partial u_p / \partial X_j$. Here, the Lagrangian $L$ that corresponds to the states with dislocations is the function of anti-symmetrical derivatives of reper components, similar to (7). The definition of Burgers vector via the reper [20] and its definition via the tensor of elastic distortion (10) are the same. This fact shows that these two approaches are related. However, the stress tensor in [20] is not defined at the value conjugated to the reper; it is introduced into the theory on the basis of additional arguments, in accordance with the Hooke law. In our opinion, if the stress tensor is not equal to zero, then it is to be contained in the equations of state as a value conjugated to the reper, because the reper is a variable and a generalization of the deformation tensor. As a result, the equations of state will be obviously changed, and the Lagrangian coefficients will be restricted by additional conditions caused by equilibrium condition. So there are two definitions of the stress tensor in the model [20], which in general case may result incompatibility of the equations of state. However, the reper was introduced in order to get the equations in which the stress tensor is not the source of dislocations, because in the region with stresses, but not with dislocations, the stresses are not the source of dislocations.

We think that this contradiction is eliminated by transition from description of finite quantity of dislocations to their density. We did not set ourselves the task of description of states with dislocations. We did set ourselves the task of description of states with local translational symmetry in the Landau theory. It was found that in these states stresses are the source of dislocations, and they cannot be moved into non-deformed states by diffeomorphism in any region of macro-space $\{\vec{X}\}$. Symmetry is an exact concept, so the linear defects appear when local regions with different periods are combined together. The density of these defects will be observed everywhere the OP has inhomogeneous distribution and does not equal to zero. The stresses and non-zero density of dislocations are caused by any inhomogeneity in the macro-space. Here, in the local region with coordinate $\vec{X}$ itself, we assume that the crystal is ideal. This is the mean of of local homogeneity principle. I.e., we do not deform the crystal by some diffeomorphism, but "glue" the locally homogeneous regions at each point by compensating fields. In such model, the dislocations are definitely caused by both the logical concept and the interpretation of mathematical expressions. The main differences between presented model and models [19], [20] are the symmetry-determined minimal interaction between OP and the distortion. There are no OP and no local translational symmetry of states in the models [19], [20].

Application of the present model with the local Landau potential may be extended if one examines the quantum-mechanical wave functions of density of state instead of OP. In fact, the quantum mechanical postulates are in agreement with the principle of local homogeneity [3]. To introduce minimal interaction, it is necessary to require the translational symmetry of the density of state inside a macroscopically small region. The quantum-mechanical concepts of matter fields that have wave properties and, hence, translation symmetry at small distances, are excellently suited for the applications of described formalism. The only assumption we shall make is that translational symmetry of the wave function is described by a continuous differentiable function. I.e., if density of state is described by a wave function with vector $\vec{k}_1$ inside some macro-volume with coordinate $\vec{X}_1$, then the vector $\vec{k}_2$ inside next volume with coordinate $\vec{X}_2$ would have a value close to $\vec{k}_1$. We think this is a reasonable assumption. It makes possible to carry the described mechanism of derivative extension over the quantum-mechanical functions of density of state. Note that in the pioneer work of Ginzburg and Landau, the function of density of state acts as an OP [1], and the model (23) constructed as a model with an OP, may be considered also as a model with a wave function. In this case, the transition to the effective potential describing the pair of opposite wave vectors, is proved.

The author wishes to acknowledge Yu. M. Gufan, for invaluable discussions and support while writing the present paper that cannot be overestimated.